\documentclass[12pt,a4paper]{article}

\usepackage[T1]{fontenc}
\usepackage[utf8]{inputenc}
\usepackage{lmodern}
\usepackage{microtype}

\usepackage[margin=2.5cm]{geometry}
\usepackage{authblk}

\usepackage{amsmath,amssymb}
\usepackage{graphicx}
\usepackage{booktabs}
\usepackage{float}
\usepackage{enumitem}

\usepackage[authoryear,longnamesfirst]{natbib}

\usepackage[colorlinks=true,linkcolor=blue,citecolor=blue,urlcolor=blue]{hyperref}
\hypersetup{
  pdftitle={The Structural Number of Parties: An Electoral-Outcome-Independent
            Baseline for Comparing Party-System Scale},
  pdfauthor={Adama Nouboukpo, Kodzo Michel Aladji, Bappa Muktar}
}

\begin{document}

\title{The Structural Number of Parties: An Electoral-Outcome-Independent
Baseline for Comparing Party-System Scale}

\author[1]{Adama Nouboukpo\thanks{Corresponding author:
\texttt{edu.nouboukpoadama@gmail.com}; \texttt{noua06@uqo.ca}}}
\author[1]{Kodzo Michel Aladji}
\author[1]{Bappa Muktar}
\affil[1]{University of Quebec at Outaouais (Universit\'e du Qu\'ebec en Outaouais),
283 Boul.\ Alexandre-Tach\'e, Gatineau, Quebec J8X 3X7, Canada}

\date{\today}
\maketitle

\begin{abstract}
How many parties can a country structurally sustain? Standard measures such as the effective number of parties (ENP) cannot answer this question. They are descriptive and post hoc: they summarize competition only after an election, and they depend on electoral returns that are often missing, interrupted, or unreliable in many developing democracies, particularly in Africa. They also offer no common scale for comparing the size of party registers with realized electoral fragmentation. We introduce the Structural Number of Parties (SNP), an outcome-independent baseline derived from two universally available quantities: population size and territorial area. Building on the logic of the seat product model and the cube-root law of assembly size, the SNP gives an ex ante benchmark that can be computed for any country and year, including where electoral data are absent. Applied to African states, the SNP enables a structural gap diagnosis that distinguishes under-fragmented systems, marked by hegemonic-party dominance, from over-fragmented ones, marked by excessive atomization. The framework thus provides a transparent and reproducible tool to inform institutional reform and the rationalization of party systems across the continent.
\end{abstract}


\section{Introduction}
The number of political parties shapes representation, government stability, and policy coordination~\citep{ricciuti2004,pildes2021}.
In many African states, particularly in Francophone countries, administrative party registers coexist with electoral competition concentrated among few actors. The Democratic Republic of the Congo (DRC) is an extreme case: 910 parties were officially registered in June 2023 \citep{ceni2023,evidencityDRCongo}, and a subsequent country report describes roughly 1{,}000 parties in 2024~\citep{bti202DRCongo}, while seats were distributed among a far smaller set of effective competitors \citep{evidencityDRCongo}.

The gap between legal recognition and effective competition has been linked to weak organization, personalism, and limited programmatic differentiation~\citep{erdmann2008,osei2006,bayart2009}. It has also prompted reform proposals that limit ``\textit{shop parties}'' driven by local or personal interests~\citep{bayart2009}, including registration thresholds and multi-regional presence requirements. Such proposals carry known risks: party regulation in Africa has repeatedly been used to exclude opposition and ethnic parties~\citep{basedau2007ethnic,moroff2010party}. Any measure intended to inform this debate therefore requires an objective framework to measure fragmentation and establish context-specific thresholds for party viability.

Political science offers rigorous measures of realized fragmentation. The Laakso--Taagepera effective number of parties (ENP) remains the principal concentration-based index~\citep{laakso1979}; Golosov's formulation addresses particular upper-tail properties~\citep{golosov2010}; and the Seat Product Model derives the expected number of parties from assembly size and district magnitude~\citep{taagepera2007}. Each answers a question about observed or institutionally expected competition. They require electoral or institutional inputs and are consequently unavailable or difficult to interpret where elections are absent, interrupted, or heavily constrained. None provides a common scale on which a party register and an effective number can be placed side by side, and the first two are unavailable where elections are absent or suspended. They provide no structural benchmark for assessing how many parties a state can sustainably support.

This article proposes such a scale. This work poses a fundamentally different question: \emph{how many parties would we expect a state of a given scale to sustain?} Answering this requires a metric fundamentally distinct from existing ENP indices. To this end, we introduce the \emph{Structural Number of Parties} (SNP), which operationalizes a core dimension of representational complexity using two universally available and comparatively stable variables such as population and area of a given country. Its value lies in two fundamental properties: it is computable for every sovereign state; it does not depend on the electoral outcomes. We introduce two ratio indicators comparing the benchmark with registered-party counts and ENP estimates and one ratio for comparing registration with effective competition.

The contribution of this work is threefold. First, we introduce for the first time the concept of the SNP, an Electoral-Outcome-Independent measure which is computable for every sovereign state, including those without competitive elections, and state its mathematical properties. Second, we introduce two benchmark-relative gaps and one benchmark-free ratio for comparing registration with effective competition. To our knowledge this pattern has not been characterised systematically, and it is invisible to either the register count or the ENP taken alone. Third, we distinguish structural scale from realized competition by comparing the SNP with three existing measures across 50 African states.

\section{Measuring Party-System Fragmentation}
\subsection{Observed and institutionally expected fragmentation}
The Laakso--Taagepera index is calculated as $N_{LT}=1/\sum_i s_i^2$, where $s_i$ is a party's vote or seat share. Golosov's index modifies how large and small parties contribute to the effective count~\citep{golosov2010}. Both measures summarize an observed distribution and are indispensable when the analytical object is electoral concentration. The Seat Product Model addresses a related but distinct problem by connecting assembly size and district magnitude to the expected effective number of parties~\citep{taagepera2007}. Because assembly size itself scales roughly with the cube root of population \citep{Taagepera1972}, the Seat Product Model already carries population information, a point that matters for using population in our baseline.

All three measures depend on electoral or institutional inputs. Electoral rules mediate how social heterogeneity becomes party fragmentation~\citep{ordeshook1994,amorimneto1997,clark2006,stoll2013changing}, and dominant-party regimes can produce low ENP in socially heterogeneous societies~\citep{bogaards2004}. In Africa, the authoritarian origins of party systems shape both concentration and institutionalization~\citep{riedl2014authoritarian,kuenzi2001party}. ENP is therefore a measure of realized competition, not of latent representational complexity.

Several African states lack usable recent results: Eritrea since 1993, Libya between 2014 and 2021, Somalia until 2021, and Mali, Burkina Faso and Niger following military takeovers in 2021--2023. Even when elections do take place, the results may reflect circumstantial balances of power rather than genuine fragmentation~\citep{koffi2021,lindberg2006}. Extensions of the ENP incorporating legislative or cabinet power \citep{blau2008} require party-level data that V-Dem \citep{vdem_dataset} and the Comparative Constitutions Project \citep{ccp_dataset} do not provide comparably across all African states.

\subsection{Registration and competition as distinct analytical objects}
Registered-party counts and the Effective Number of Parties (ENP) capture fundamentally different dimensions of party systems. The former measures legal or administrative entry, while the latter captures the concentration of votes or seats. Party registers may include inactive organizations, personalistic vehicles, and parties lacking durable territorial or programmatic foundations. Conversely, a low ENP may reflect voter coordination, restrictive institutions, or incumbent dominance. The two measures are therefore complementary rather than interchangeable. Their divergence is itself substantively important because it indicates how much formal pluralism translates into effective competition.

Cameroon illustrates the divergence. The Cameroon People's Democratic Movement has kept seat-based ENP below 2 across successive elections \citep{bti2026Camer}, while the register lists approximately 300 parties. \citet{bogaards2004} shows that in such cases low ENP reflects dominance rather than social simplicity. This distortion is driven by the proliferation of ``\textit{shop parties}'' (formations lacking territorial presence or programmatic content, often serving narrow local or personal interests), which artificially inflate registration counts while leaving effective electoral fragmentation unchanged.

\subsection{The missing common denominator}
A register count and an effective number arise from different processes, so their raw values cannot be meaningfully compared on one scale. The DRC's Golosov ENP of 36.67 and its 910 registered formations are not commensurable as they stand. A common denominator cannot explain either quantity, but it can express both relative to the same reference. The SNP is designed for that limited purpose.

\section{The Structural Number of Parties}
The SNP and ENP answer different questions. The shift from the ENP to the SNP is paradigmatic in the sense of \citet{kuhn1962}: it changes the question rather than refining the answer. Where ENP asks how many parties have emerged from a given electoral context, SNP asks how many parties a state can structurally sustain.

Table~\ref{tab:paradigm} summarizes this distinction. The two approaches are complementary: confronted with the SNP, the ENP becomes interpretable as a measure of deviation from structural capacity. We formalise this deviation through the ratio indicators, in the next section, thereby translating the paradigm shift into a measurable diagnostic tool.

\begin{table}[H]
\centering
\caption{Paradigmatic comparison of ENP and SNP.}
\label{tab:paradigm}
\scalebox{0.85}{%
\begin{tabular}{|p{3.8cm}|p{5.2cm}|p{5.2cm}|}
\hline
\textbf{Dimension} & \textbf{ENP} & \textbf{SNP} \\ \hline
Question & How many parties have effective influence after an election? & How many parties should a state structurally support? \\ \hline
Nature & Descriptive and retrospective & Prescriptive and structural \\ \hline
Data requirements & Electoral results (votes, seats) & Population, territorial area \\ \hline
Availability & Impossible in the absence of elections & Available for any sovereign state \\ \hline
Institutional sensitivity & High (depends on the electoral system) & None (independent of the electoral system) \\ \hline
Validity in the African context & Partial and conditional & Broad cross-national availability \\ \hline
\end{tabular}}
\end{table}

\subsection{Why population and territory}
The SNP is grounded in two structural attributes: population size and territorial area. Both variables are widely available, comparatively stable, and less directly shaped by short-term electoral outcomes than institutional measures such as district magnitude or electoral-system design. They are not strictly exogenous or error-free: population estimates may be affected by census quality, migration, conflict, and demographic shocks, while territorial boundaries are historically and politically constituted. Nevertheless, relative to electoral and party-system indicators, population and area provide slow-moving and broadly comparable measures of state scale. They are used here to build a parsimonious scale, not to identify causes of fragmentation.

Population approximates the scale of potential representational demand. Larger populations can contain more interests and identities and offer parties larger constituencies \citep{dahl1973}. \citet{sikk2014population} show that population carries systematic information about the effective number of legislative parties, although the relationship varies with electoral systems \citep{amorimneto1997}.

Territorial area captures spatial dispersion and the organizational cost of national coordination \citep{herbst2000}. Its expected effect on the number of parties is theoretically ambiguous. Dispersion can favour regional parties where decentralization creates subnational arenas \citep{brancati2008regional}, but high coordination costs and centralized authority can push candidates toward national labels and reduce the number of viable parties \citep{chhibber1998aggregation,chhibber2004formation}. Ethnopolitical geography mediates both channels in Africa \citep{mozaffar2003}. Territory should therefore be understood not as a direct measure of social diversity, but as the spatial setting across which political identities, institutions, and organizations interact. We include it because a reference scale for party systems should register that a large, sparse state and a small, dense one of equal population pose different organizational settings. Whether area adds information beyond population is an empirical question we address in the results section.

Population and territorial area thus capture complementary dimensions of structural political complexity. Population approximates the scale of potential representational demand, whereas territory approximates its spatial dispersion and the organizational costs of national integration. A populous but geographically compact state may combine substantial social complexity with relatively limited coordination costs. Conversely, a sparsely populated but territorially extensive state may confront significant barriers to nationwide organization despite its smaller electorate. Combining the two variables provides a parsimonious measure of demographic and territorial scale while deliberately abstracting from endogenous electoral outcomes and institutional arrangements.

\subsection{Specification and scale convention}
For a country with population $P>0$, measured in inhabitants, and territorial area $A>0$, measured in square kilometres, let $N_{\mathrm{SNP}}$ denote the SNP. The Structural Number of Parties is defined as:
\begin{equation}
N_{\mathrm{SNP}}
=\max\left\{
2,\;
\ln\left[
\left(\frac{P}{P_0}\right)
\left(\frac{A}{A_0}\right)
\right]
\right\},
\label{eq:snp}
\end{equation}
where $P_0$ and $A_0$ serve as normalization constants, ensuring the argument of the logarithm is dimensionless and establishing a well-defined numerical scale.

In this work, we set $P_0=10^6$ inhabitants and $A_0=10^4\,\mathrm{km}^2$, so $C=P_0A_0=10^{10}$ in the stated units (more details in Appendix~\ref{a0poexplain}). The $\max$ operator imposes a lower bound of 2, reflecting the minimal condition for meaningful party competition: at least two viable alternatives must exist for electoral choice to be non-trivial. A system with fewer than two effective parties is, by definition, hegemonic or non-competitive. This floor is not arbitrary but grounded in the theoretical distinction between competitive and hegemonic party systems \citep{sartori1976}.

\subsection{Properties}
The specification has five properties (proofs in Appendix~\ref{snpexplain}):
\begin{enumerate}[leftmargin=*,itemsep=4pt]
\item \textbf{Democratic floor:} $N_{\mathrm{SNP}}\geq 2$.
\item \textbf{Monotonicity in demographic and territorial scale:}
Holding territorial area constant, an increase in population cannot reduce the SNP; similarly, holding population constant, an increase in territorial area cannot reduce it.
\item \textbf{Decreasing marginal effects:} above the floor, the marginal contribution of either input declines as that input grows.
\item \textbf{Equal proportional sensitivity:} the same proportional change in population or area produces the same change in the uncensored SNP.
\item \textbf{Electoral-outcome independence:} the calculation uses no votes, seats, district magnitude, party count, or regime classification.
\end{enumerate}
Property 5 marks the boundary between the SNP and ENP measures.

\subsection{Diagnosing fragmentation through structural gaps}
Let $N_f$ denote the number of legally registered parties and $N_G$ the Golosov ENP. Taking $N_{\mathrm{SNP}}$ as the structural benchmark, we define three gaps:
\begin{equation}
\rho_f=\ln\left(\frac{N_f}{N_{\mathrm{SNP}}}\right),\qquad
\rho_e=\ln\left(\frac{N_G}{N_{\mathrm{SNP}}}\right),\qquad
\delta=\ln\left(\frac{N_f}{N_G}\right)=\rho_f-\rho_e.
\label{eq:gaps}
\end{equation}
The formal-registration gap, $\rho_f$, measures the size of the party register relative to the structural benchmark, whereas the effective-competition gap, $\rho_e$, measures effective electoral fragmentation relative to the same benchmark. Both indicators therefore express observed party-system quantities relative to $N_{\mathrm{SNP}}$. Positive values indicate that the observed quantity exceeds the benchmark; negative values indicate that it falls below it; and zero denotes equality. The discrepancy $\delta$ does not: $N_{\mathrm{SNP}}$ cancels, and $e^\delta=N_f/N_G$ is the number of registered parties per effective party. A larger value indicates a greater discrepancy between formal party registration and effective fragmentation.

We emphasize the division of labour. $\delta$ requires no SNP and is reported for completeness. The SNP's contribution is the location of each quantity on a common scale, which allows cases with similar $\delta$ but very different competitive structures to be told apart (Section~\ref{gapanalysis}). Because that location depends on calibration, Section~\ref{calibsens} reports how far it moves when the reference changes.

\section{Data and Empirical Strategy}
The analysis uses three nested samples: all 54 African United Nations member states for the SNP itself, 50 states for the comparison with existing measures, and 17 states for the registration--competition diagnosis.

\paragraph{Inputs.} Population and area come from the World Bank World Development Indicators (WDI)~\citep{worldbank2025}, merged on ISO country codes to avoid confusion between the two Congos and among Guinea, Guinea-Bissau and Equatorial Guinea. Population is measured in inhabitants and territorial area in square kilometres. Both inputs are positive and non-missing for all 54 states, with no duplicates. The resulting dataset contains country names, country codes, annual population estimates, and territorial area. All calculations use unrounded SNP values; values presented in tables and figures are rounded for readability.

\paragraph{Temporal stability.} A structural baseline should remain comparatively stable over short periods and change only when its underlying demographic or territorial inputs change. Because territorial area is largely static, we assess the SNP's temporal stability by computing it annually from 2013 to 2023, holding area constant and updating population for each year.

\paragraph{Registration--competition diagnosis.}
The registration--competition analysis covers the 17 states for which both a registered-party count and a Golosov ENP estimate could be compiled. For each country, we calculate the formal-registration gap, the effective-competition gap, and the discrepancy gap between registered and effective parties. Registration counts and electoral observations do not necessarily refer to the same year. The analysis should therefore be interpreted as a cross-sectional diagnostic of the relationship between formal party registration and realized electoral competition, rather than as a contemporaneous country ranking or a causal design. The replication materials report, for each observation, the registration date, election year, legislative chamber, vote or seat shares, treatment of coalitions and independent candidates, and original data source.

\paragraph{Calibration sensitivity.}
Because $\rho_f$ and $\rho_e$ depend on the reference product $C$, we recompute them with $C$ multiplied by $k \in \{e^{-1}, e^{-0.5}, e^{0.5}, e^{1}\}$, which shifts every unfloored score by $-\ln k$ (Appendix~\ref{refchange}). We report the number of cases whose sign classification changes and, for each case, the shift at which $\rho_e$ would change sign.

\paragraph{Comparison with existing measures.}
To examine how the SNP relates to established measures of party-system size, we compare it with the Laakso--Taagepera ENP, the Golosov ENP, and the Seat Product estimate. For the 50 states with available estimates, the comparison is evaluated using Spearman rank correlations and bivariate ordinary least-squares (OLS) regressions. Spearman's coefficient assesses whether the measures produce similar country rankings without assuming a linear relationship, whereas the regressions summarize their linear association through the estimated slope and explained variance. We additionally report a leave-one-out estimate excluding DRC, a coalition-aggregated alternative for DRC, a joint regression of ENP on the SNP and the Seat Product estimate, and a comparison of the SNP with $\ln(\text{population})$ alone. Eritrea, Eswatini, Libya, and Somalia are excluded due to the absence of comparable electoral data required to compute the ENP.

\section{Results}
\subsection{Structural scale across Africa and temporal stability}
Table~\ref{tab:snp54} presents the distribution of the \emph{SNP} index across African countries. The SNP ranges from 2.00 to 10.12, with a median of 6.06 and an interquartile range of 3.77--7.87. DR Congo records the highest value (10.12), followed by Nigeria (9.95), Ethiopia (9.56), Egypt (9.35), and Algeria (9.31). Ten states, mostly small island and enclave states, sit at the floor of 2.00. Area weighs heavily for large, sparsely populated states. Libya (7.16) and Mauritania (6.25) score close to Ghana (6.69) and C\^ote d'Ivoire (6.91), despite much smaller populations. Furthermore, because the SNP calculation relies exclusively on population and land area, it provides a reliable baseline even for states with missing, unreliable, or non-existent electoral data (e.g., Eritrea, Libya, Somalia, and Eswatini).

\begin{table}[H]
\centering
\small
\caption{Structural Number of Parties in African States.}
\label{tab:snp54}
\begin{tabular}{@{}lr@{\qquad}lr@{\qquad}lr@{}}
\toprule
\textbf{Country} & \textbf{SNP} &
\textbf{Country} & \textbf{SNP} &
\textbf{Country} & \textbf{SNP} \\
\midrule
Algeria                & 9.31  & Eswatini          & 2.00 & Namibia               & 5.50 \\
Angola                 & 8.43  & Ethiopia          & 9.56 & Niger                 & 8.11 \\
Benin                  & 5.09  & Gabon             & 4.20 & Nigeria               & 9.95 \\
Botswana               & 4.97  & Gambia            & 2.00 & Rwanda                & 3.60 \\
Burkina Faso           & 6.45  & Ghana             & 6.69 & Sao Tome \& Principe  & 2.00 \\
Burundi                & 3.64  & Guinea            & 5.87 & Senegal               & 5.87 \\
Cameroon               & 7.21  & Guinea-Bissau     & 2.05 & Seychelles            & 2.00 \\
Cape Verde             & 2.00  & Kenya             & 8.07 & Sierra Leone          & 4.11 \\
Central African Rep.   & 5.77  & Lesotho           & 2.00 & Somalia               & 7.07 \\
Chad                   & 7.82  & Liberia           & 4.11 & South Africa          & 8.95 \\
Comoros                & 2.00  & Libya             & 7.16 & South Sudan           & 6.61 \\
Congo                  & 5.35  & Madagascar        & 7.51 & Sudan                 & 9.14 \\
C\^ote d'Ivoire        & 6.91  & Malawi            & 5.52 & Tanzania              & 8.75 \\
DR Congo               & 10.12 & Mali              & 7.99 & Togo                  & 3.97 \\
Djibouti               & 2.00  & Mauritania        & 6.25 & Tunisia               & 5.30 \\
Egypt                  & 9.35  & Mauritius         & 2.00 & Uganda                & 7.07 \\
Equatorial Guinea      & 2.00  & Morocco           & 7.89 & Zambia                & 7.35 \\
Eritrea                & 3.71  & Mozambique        & 7.90 & Zimbabwe              & 6.46 \\
\bottomrule
\end{tabular}
\end{table}

\begin{figure}[H]
\centering
\includegraphics[width=1.0\linewidth]{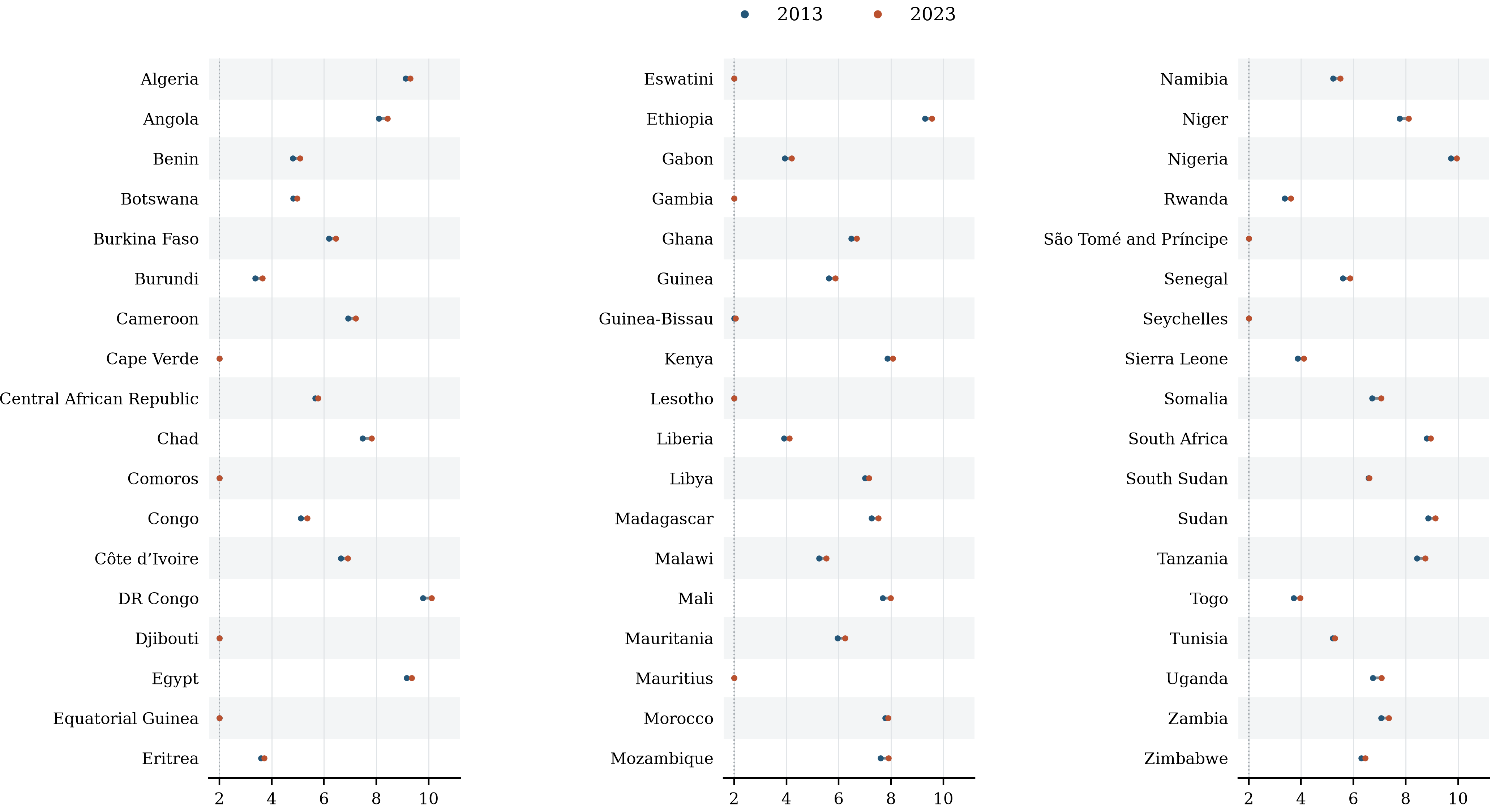}
\caption{Temporal stability of the SNP, 2013--2023.}
\label{fig:trends}
\end{figure}

Figure~\ref{fig:trends} shows that annual changes are modest relative to cross-sectional differences. Between 2013 and 2023, the median country-level increase in the SNP was 0.231 points and the mean 0.191; the largest was 0.355 (Somalia). The mean falls below the median because the ten floored states record no change. Under the fixed-area specification, each change equals the log of population growth (Appendix~\ref{app:A3}), so a decade of rapid demographic growth moves the reference by about a third of a point.

\subsection{Registration--competition diagnosis}\label{gapanalysis}
Table~\ref{tab:diagnosis} separates three comparisons: registration relative to the SNP, effective fragmentation relative to the SNP, and registration relative to effective fragmentation. All 17 registered-party counts exceed the structural reference. The median formal-registration gap is $\rho_f=2.66$, corresponding to a register approximately 14.29 times the SNP. By contrast, Golosov ENP lies below the reference in 14 countries. Its median gap is $\rho_e=-0.87$, equivalent to an effective-party count approximately 0.42 times the benchmark. Large registers thus generally coexist with effective fragmentation below the same reference. Cameroon and Senegal combine particularly large registration gaps with negative effective-competition gaps. DR Congo occupies a different position: both its registered count and its Golosov ENP substantially exceed the SNP. Guinea-Bissau also has a positive effective gap, whereas the Central African Republic is effectively at parity. Its unrounded $\rho_e$ is approximately 0.00494 and rounds to 0.00 in the table; its position just above zero should not be treated as a substantively distinct category.

The discrepancy (benchmark-free) ratio, $e^\delta=N_f/N_G$, ranges from 3.21 in Benin to 218.98 in Cameroon, with a median of 25.00. It exceeds ten in 13 countries and equals ten in Equatorial Guinea; Senegal records the second-largest ratio, at 158.20. These quantities express registered-party counts as multiples of an effective-number index. Similar ratios can also conceal different competitive structures: Guinea and DR Congo both have ratios close to 25, despite Golosov ENP values of 1.60 and 36.67. Reading the direct ratio alongside the benchmark-dependent gaps preserves this distinction.

\begin{table}[H]
\centering
\footnotesize
\caption{Registration--competition diagnosis in 17 African states.}
\label{tab:diagnosis}
\begin{tabular}{lrrrrrrr}
\toprule
Country & SNP & $N_f$ & $N_G$ & $\rho_f$ & $\rho_e$ & $\delta$ & $e^\delta$ \\
\midrule
Benin                 & 5.09  & 10  & 3.12  & 0.68 & $-0.49$ & 1.16 & 3.21 \\
Burundi               & 3.64  & 35  & 1.45  & 2.26 & $-0.92$ & 3.18 & 24.14 \\
Cameroon              & 7.21  & 300 & 1.37  & 3.73 & $-1.66$ & 5.39 & 218.98 \\
Comoros               & 2.00  & 15  & 1.22  & 2.01 & $-0.49$ & 2.51 & 12.30 \\
Congo                 & 5.35  & 70  & 1.45  & 2.57 & $-1.31$ & 3.88 & 48.28 \\
C\^ote d'Ivoire       & 6.91  & 150 & 2.46  & 3.08 & $-1.03$ & 4.11 & 60.98 \\
Djibouti              & 2.00  & 8   & 1.07  & 1.39 & $-0.63$ & 2.01 & 7.48 \\
Gabon                 & 4.20  & 60  & 1.62  & 2.66 & $-0.95$ & 3.61 & 37.04 \\
Guinea                & 5.87  & 40  & 1.60  & 1.92 & $-1.30$ & 3.22 & 25.00 \\
Equatorial Guinea     & 2.00  & 10  & 1.00  & 1.61 & $-0.69$ & 2.30 & 10.00 \\
Guinea-Bissau         & 2.05  & 30  & 3.35  & 2.68 & 0.49    & 2.19 & 8.96 \\
Madagascar            & 7.51  & 200 & 2.72  & 3.28 & $-1.02$ & 4.30 & 73.53 \\
Central African Rep.  & 5.77  & 80  & 5.80  & 2.63 & 0.00    & 2.62 & 13.79 \\
DR Congo              & 10.12 & 910 & 36.67 & 4.50 & 1.29    & 3.21 & 24.82 \\
Senegal               & 5.87  & 386 & 2.44  & 4.19 & $-0.88$ & 5.06 & 158.20 \\
Chad                  & 7.82  & 150 & 3.26  & 2.95 & $-0.87$ & 3.83 & 46.01 \\
Togo                  & 3.97  & 118 & 1.67  & 3.39 & $-0.87$ & 4.26 & 70.66 \\
\midrule
Median                &       &     &       & 2.66 & $-0.87$ & 3.22 & 25.00 \\
\bottomrule
\end{tabular}
\begin{flushleft}
\footnotesize Notes: $N_f$ is the registered-party count and $N_G$ is Golosov ENP. Rounded values may not reproduce the displayed gaps exactly.
\end{flushleft}
\end{table}

\begin{figure}[H]
\centering
\includegraphics[width=0.98\linewidth]{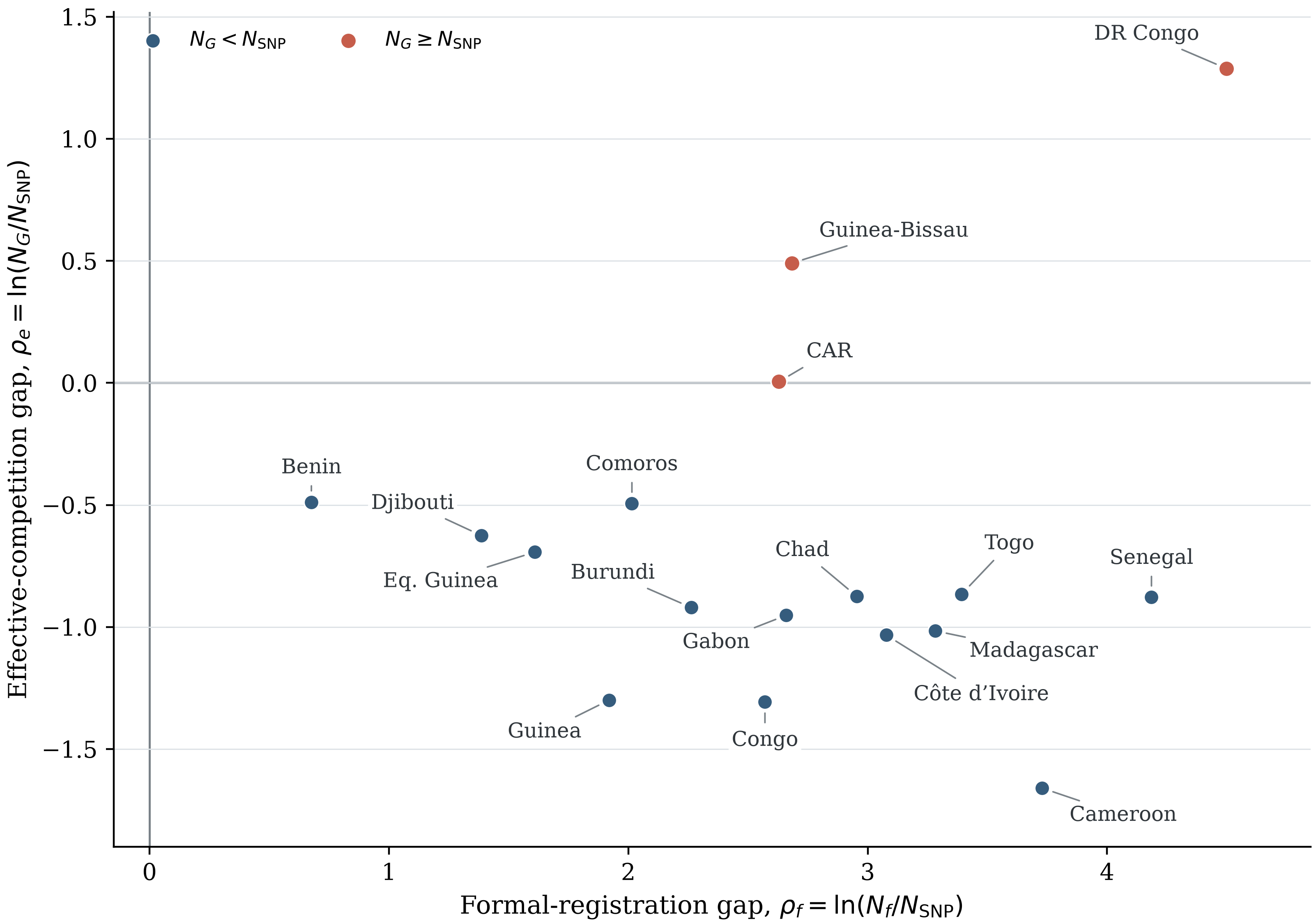}
\caption{Formal-registration and effective-competition gaps. Positive $\rho_f$ indicates a register above the SNP; positive $\rho_e$ indicates Golosov ENP above the SNP.}
\label{fig:diagnosis}
\end{figure}

We can observe two patterns: the first is consistent with a dominant party facing a crowded register; the second with genuine fragmentation alongside it. Figure~\ref{fig:diagnosis} makes this joint pattern visible. All observations lie to the right of the zero-registration line, while most lie below the zero-competition line.

\subsection{Calibration sensitivity}\label{calibsens}
The sign classifications in Table~\ref{tab:diagnosis} are robust to the reference constant for 16 of 17 states. Varying $C$ by factors between $e^{-1}$ and $e^{1}$ shifts every unfloored SNP by up to one point in either direction. Across that range all 17 registers remain above the reference, and the number of cases with ENP below it moves only between 14 and 15 (Table~\ref{tab:sens}). The scores are recomputed from unrounded, uncensored values, so floor effects are handled exactly.

Only the Central African Republic changes sign, because its ENP almost equals its SNP. Guinea-Bissau's positive gap survives increases in the reference of approximately 1.3 points; DR Congo's survives any plausible recalibration.

\begin{table}[htbp]
\centering
\caption{Gap classifications under alternative reference products $C' = kC$ ($n = 17$).}
\label{tab:sens}
\small
\begin{tabular}{lrrrrr}
\toprule
$k$ & SNP shift & Registers above SNP & ENP below SNP & Median $\rho_f$ & Median $\rho_e$ \\
\midrule
$e^{1}$        & $-1.0$ & 17 & 14 & 2.82 & $-0.68$ \\
$e^{0.5}$      & $-0.5$ & 17 & 14 & 2.72 & $-0.77$ \\
1 (baseline)   & 0      & 17 & 14 & 2.66 & $-0.87$ \\
$e^{-0.5}$     & $+0.5$ & 17 & 15 & 2.55 & $-0.96$ \\
$e^{-1}$       & $+1.0$ & 17 & 15 & 2.45 & $-1.04$ \\
\bottomrule
\end{tabular}
\vspace{0.5em}
\begin{minipage}{0.92\textwidth}
\footnotesize Notes: computed from unrounded, uncensored scores, $N'_{\mathrm{SNP}} = \max\{2,\, g - \ln k\}$ (Appendix~\ref{refchange}). The only sign change is the Central African Republic, at $k \le e^{-0.5}$.
\end{minipage}
\end{table}

This robustness has two sources that should be distinguished. For nine of the 14 negative cases (Burundi, Cameroon, Comoros, Congo, Djibouti, Equatorial Guinea, Gabon, Guinea and Togo), Golosov ENP is below 2. Because the SNP can never fall below 2, their negative competition gap holds under every calibration: it follows from the floor convention and identifies dominant-party outcomes, not a finding about scale. For the other five (Benin, Chad, C\^ote d'Ivoire, Madagascar and Senegal), ENP lies between 2 and the reference, and the sign would change only if the reference fell by 1.97 points (Benin) to 4.79 points (Madagascar). These five cases, together with the three positive ones, are where the SNP's scale carries information. Conclusions about the direction of the gaps therefore do not hinge on the choice of $C$; their magnitudes do, and we interpret them only comparatively.

\subsection{Correspondence with existing measures}
The SNP and conventional party-system indices capture different analytical objects. Their comparison is therefore not intended to assess predictive accuracy or determine which measure is superior. Instead, it examines whether the SNP captures a dimension of country scale that is empirically distinct from realized electoral competition. The SNP is nearly unrelated to realized fragmentation and closely related to the Seat Product estimate (Table~\ref{tab:enp-associations}).

Table~\ref{tab:enp-associations} reports associations between the SNP and three existing measures for the 50 countries with available observations. The SNP is weakly and nonsignificantly associated with both Laakso--Taagepera ENP ($\rho=0.083$, $p=0.565$) and Golosov ENP ($\rho=0.061$, $p=0.673$). The corresponding bivariate regressions explain only 5.7 and 5.1 per cent of the variation, respectively. These weak relationships are consistent with the conceptual distinction between structural country scale and realized electoral concentration. They do not, however, constitute independent evidence of the SNP's validity.

\begin{table}[htbp]
\centering
\caption{Associations between the SNP and existing measures across African states.}
\label{tab:enp-associations}
\small
\begin{tabular}{lrrrrrr}
\toprule
Measure & $n$ & Spearman $\rho$ & $p$ & OLS slope & $p$ & $R^2$ \\
\midrule
Laakso--Taagepera ENP & 50 & 0.083 & 0.565 & 0.435 & 0.096 & 0.057 \\
Golosov ENP & 50 & 0.061 & 0.673 & 0.445 & 0.115 & 0.051 \\
Seat Product estimate & 50 & 0.889 & $<0.001$ & 0.455 & $<0.001$ & 0.731 \\
Laakso--Taagepera ENP, excl.\ DR Congo & 49 & 0.026 & 0.860 & 0.020 & 0.827 & 0.001 \\
Golosov ENP, excl.\ DR Congo & 49 & 0.002 & 0.988 & $-0.013$ & 0.876 & 0.001 \\
Seat Product estimate, excl.\ DR Congo & 49 & 0.884 & $<0.001$ & 0.450 & $<0.001$ & 0.715 \\
\bottomrule
\end{tabular}
\vspace{0.5em}
\begin{minipage}{0.92\textwidth}
\footnotesize Notes: OLS regresses each measure on the SNP.
\end{minipage}
\end{table}

\begin{figure}[htbp]
\centering
\includegraphics[width=0.98\linewidth]{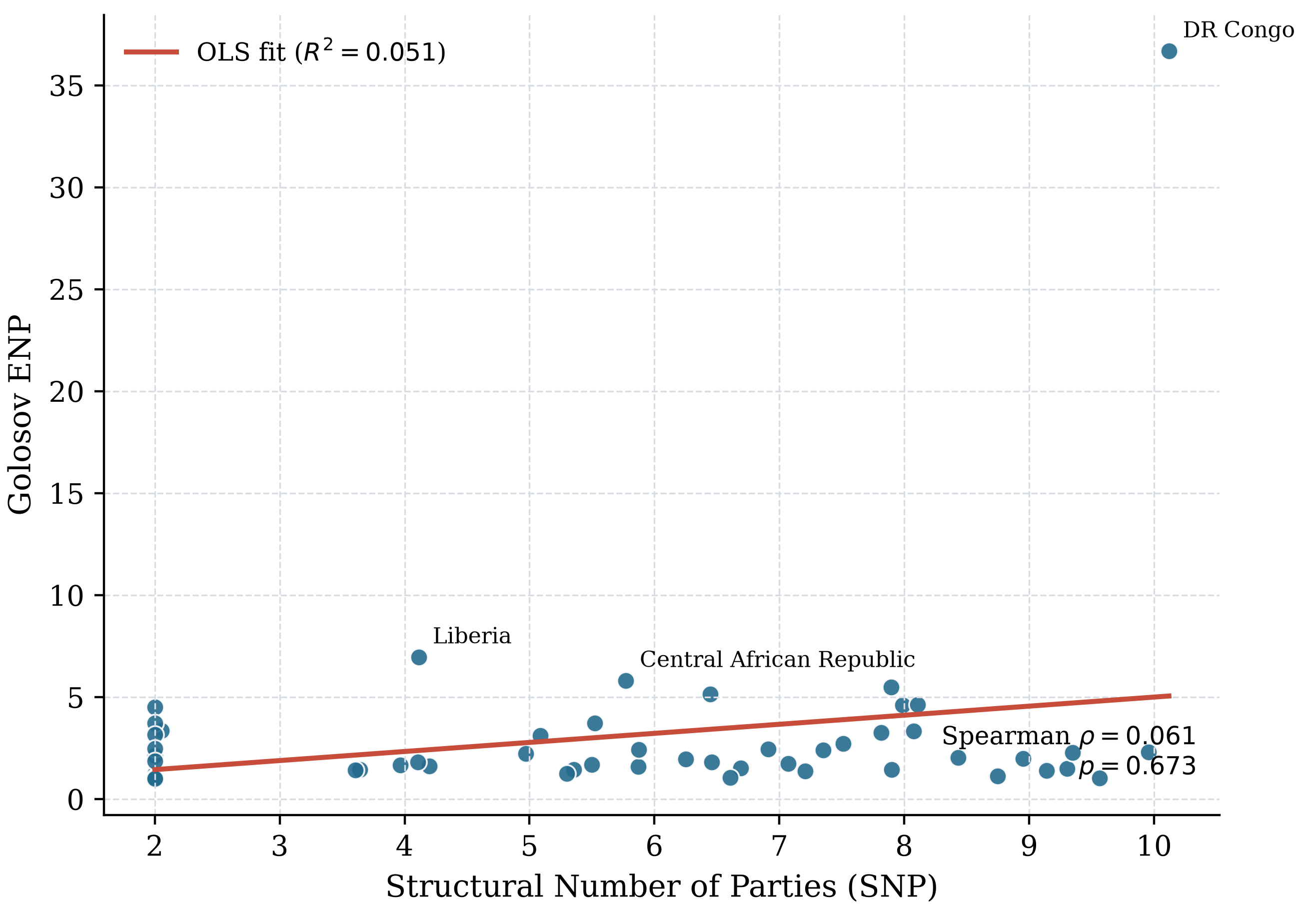}
\caption{Structural Number of Parties and Golosov ENP across 50 African states. The fitted line is a bivariate ordinary least-squares estimate.}
\label{fig:golosov}
\end{figure}

The gap between the rank and linear results for ENP reflects DR Congo, which combines the highest SNP with an ENP of 36.67. Excluding it, the OLS slope for Golosov ENP falls from 0.445 to $-0.013$ and $R^2$ from 0.051 to 0.001; the Laakso--Taagepera results move the same way. The weak positive linear association in the full sample is therefore entirely attributable to one case, and there is no association between structural scale and realized fragmentation among the remaining 49 states. The Seat Product association is unaffected ($\rho = 0.884$, $R^2 = 0.715$). Countries of similar scale realize very different levels of fragmentation, consistent with the mediating role of electoral rules, regime type and cleavage structure (Figure~\ref{fig:golosov}).

The strong association with the Seat Product estimate is expected rather than independently confirming. Assembly size scales with population \citep{Taagepera1972}, and the Seat Product Model is log-linear in assembly size, so both measures share a population component. Table~\ref{tab:poparea} confirms this. Population alone ranks countries almost exactly as the SNP does against the Seat Product estimate (Spearman $\rho = 0.884$ against 0.889). In a regression of the Seat Product estimate on $\ln P$ and $\ln A$ separately, the area coefficient is small and not significant (0.089, $p = 0.318$), and dropping area lowers $R^2$ only from 0.707 to 0.700.

Two conclusions follow. First, in this sample the area component adds little measurable information; a population-only reference would order countries similarly. We retain area for the conceptual reason given earlier, that spatial dispersion is part of what a scale measure should describe, but we flag that this choice is not supported by the present correspondence evidence. Second, neither the SNP nor the Seat Product estimate predicts realized ENP: in joint regressions of Golosov ENP on both, neither coefficient is significant, with or without DR Congo (all $p > 0.25$; replication table~\ref{tab:enp-associations}).
Overall, the comparison supports treating the SNP as a complementary structural denominator rather than as a substitute for election-based measures.

\begin{table}[htbp]
\centering
\caption{Population, area and the SNP as correlates of existing measures ($n = 50$).}
\label{tab:poparea}
\small
\begin{tabular}{lrrrrrrr}
\toprule
& \multicolumn{3}{c}{Spearman $\rho$ with} & \multicolumn{4}{c}{OLS on $\ln P$ and $\ln A$} \\
\cmidrule(lr){2-4}\cmidrule(lr){5-8}
Measure & $\ln P$ & $\ln A$ & SNP & $\ln A$ coef. & $\ln A$ $p$ & $R^2$ & $R^2$, $\ln P$ only \\
\midrule
Laakso--Taagepera ENP & 0.052 & 0.109 & 0.083 & 0.239 & 0.665 & 0.042 & 0.038 \\
Golosov ENP & 0.033 & 0.075 & 0.061 & 0.190 & 0.750 & 0.036 & 0.034 \\
Seat Product estimate & 0.884 & 0.758 & 0.889 & 0.089 & 0.318 & 0.707 & 0.700 \\
\bottomrule
\end{tabular}
\end{table}

\section{Conclusion}
This article introduces the \emph{Structural Number of Parties}, an apolitical and institutionally neutral index that operates independently of electoral outcomes. It addresses a fundamental question overlooked by traditional metrics: how many parties can a state structurally sustain? Grounded in two universal determinants, population and land area, the SNP eliminates reliance on electoral data, enabling complete coverage of African states even in the absence of reliable elections. Unlike \emph{ex post} approaches, the SNP provides an \emph{ex ante} normative benchmark to evaluate party system viability. Empirically, our analysis reveals significant cross-national variation in the SNP across Africa, underscoring the profound impact of geographical and demographic factors on political fragmentation. Furthermore, by introducing structural gaps (e.g., registered versus effective parties), we can differentiate between the dynamics of formal (administrative) and actual (electoral) fragmentation. The SNP thus serves as a normative benchmark for aligning party systems with their structural capacity, allowing researchers to distinguish between artificial fragmentation, institutional constraints, and structural alignment.

This study establishes a robust foundation for debates concerning the rationalization of political parties in Africa. Crucially, our findings suggest a paradigm shift in party-system reform recommendations. While this paper does not prescribe the specific mechanisms for this reduction or dictate which parties should endure, it opens new avenues for exploring the political implications of identifying an optimal number of parties for each country. Future research will investigate the practical mechanisms for achieving this rationalization, including modeling merger and cooperation scenarios among existing parties. Ultimately, fostering consolidation among ideologically aligned parties can yield more rationalized political landscapes and foster stable, effective democratic systems. Finally, although this study focuses on the African context, the SNP framework is highly adaptable; it can be applied to other global regions by recalibrating model parameters and incorporating context-specific variables.

\section*{Author Contributions}
\noindent\textbf{Adama Nouboukpo:} Conceptualization, Methodology, Resources, Software, Validation, Visualization, Investigation, Writing -- original draft.\\
\textbf{Kodzo Michel Aladji:} Resources, Writing -- review and editing.\\
\textbf{Bappa Muktar:} Writing -- review and editing.


\section*{Declaration of competing interests}
The authors declared no potential conflicts of interest with respect to the research, authorship, and/or publication of this article.

\section*{Funding}
The authors received no financial support for the research, authorship, and/or publication of this article.

\section*{Acknowledgments}
The authors would like to acknowledge the University of Quebec at Outaouais / Universit\'e du Qu\'ebec en Outaouais (UQO) for its institutional support.

\bibliographystyle{plainnat}
\bibliography{references}

\appendix

\section{Mathematical Properties of the SNP}\label{snpexplain}
For population $P>0$ and territorial area $A>0$, the SNP is given by Equation~\eqref{eq:snp}, with $P_0 = 10^6$ inhabitants and $A_0 = 10^4$ km$^2$:
\begin{equation}
N_{\mathrm{SNP}}(P,A)
=\max\left\{2,\;\ln\left[\left(\frac{P}{P_0}\right)
\left(\frac{A}{A_0}\right)\right]\right\},
\label{eq:snp-reduced}
\end{equation}
where $P_0=10^6$ inhabitants and $A_0=10^4\,\mathrm{km}^2$. Population and area are expressed in inhabitants and square kilometres, respectively, and $\ln$ denotes the natural logarithm. The ratios make its argument dimensionless. The reference quantities and lower bound are fixed conventions; no sample median, median absolute deviation, or estimated scaling parameter enters the calculation.

\subsection{Equivalent representation and sample independence}\label{a0poexplain}
Define the unfloored component as
\begin{equation}
g(P,A)=\ln\left(\frac{PA}{P_0A_0}\right)
=\ln\left(\frac{P}{P_0}\right)+\ln\left(\frac{A}{A_0}\right)
=\ln\frac{PA}{C}, \qquad C = P_0A_0 = 10^{10}.
\label{eq:snp-uncensored}
\end{equation}
Only the reference product $C$ affects the score; any reference pair with the same product gives identical results. Each score depends only on that country's inputs, so adding or removing countries leaves it unchanged.

\subsection{Monotonicity and marginal effects}
Above the floor, where $g(P,A)>2$, the first and second partial derivatives are
\begin{equation}
\frac{\partial N_{\mathrm{SNP}}}{\partial P}=\frac{1}{P}>0, \qquad
\frac{\partial N_{\mathrm{SNP}}}{\partial A}=\frac{1}{A}>0,
\end{equation}
\begin{equation}
\frac{\partial^2 N_{\mathrm{SNP}}}{\partial P^2}=-\frac{1}{P^2}<0,
\qquad
\frac{\partial^2 N_{\mathrm{SNP}}}{\partial A^2}=-\frac{1}{A^2}<0,
\qquad
\frac{\partial^2 N_{\mathrm{SNP}}}{\partial P\,\partial A}=0.
\end{equation}
Below the floor both first partials are zero; at the threshold the function is continuous but not differentiable. Globally, the index is non-decreasing in both inputs but not concave.

\subsection{Proportional changes and temporal comparisons}\label{app:A3}
For any $q>0$, the unfloored component satisfies
\begin{equation}
g(qP,A)-g(P,A)=g(P,qA)-g(P,A)=\ln q.
\end{equation}
Equal proportional changes in either input therefore have equal additive effects on $g$. This identity also holds for changes in the reported SNP when both compared observations are at or above the floor. If both remain below the threshold, the reported change is zero; when an observation crosses the threshold, the floor modifies the change.

More generally, for two observations with positive inputs,
\begin{equation}
g(P_2,A_2)-g(P_1,A_1)=\ln\left(\frac{P_2}{P_1}\right)+\ln\left(\frac{A_2}{A_1}\right).
\end{equation}
With territorial area fixed and both observations at or above the floor, this reduces to
\begin{equation}
N_{\mathrm{SNP},2}-N_{\mathrm{SNP},1}=\ln\left(\frac{P_2}{P_1}\right).
\end{equation}
So temporal change reflects population growth only. The logarithmic transformation moderates absolute score changes, while the floor can produce unchanged scores despite changing inputs.

\subsection{Lower bound and ordering}
The floor binds whenever
\begin{equation}
\frac{PA}{C}\leq e^2.
\end{equation}
Above it, the ordering of SNP values equals the ordering of the population--area product. The index has no upper bound.

\subsection{Reference changes and diagnostic gaps}\label{refchange}
Let an alternative reference product be $C'=kC$, with $k>0$, holding the measurement units fixed. Its unfloored score is
\begin{equation}
g'(P,A)=g(P,A)-\ln k, \qquad
N'_{\mathrm{SNP}}=\max\{2,\;g(P,A)-\ln k\}.
\end{equation}
The ordering of unfloored scores is preserved, but floor incidence and ties can change. For any positive alternative benchmark,
\begin{equation}
\rho'_f-\rho_f=\rho'_e-\rho_e
=\ln\left(\frac{N_{\mathrm{SNP}}}{N'_{\mathrm{SNP}}}\right),
\qquad
\delta'=\delta.
\end{equation}
Benchmark-relative gaps shift equally within a country, although the shift can differ across countries when some are floor-bound; $\delta$ and $e^\delta = N_f/N_G$ are invariant.

\end{document}